\documentstyle[preprint,aps]{revtex}
\begin{document}
\tighten
\preprint{\vbox{\noindent
\hfill {\it published in:} Phys. Rev. C Rapid Comm.\\
\null\hfill vol. 58(6), p. R3055 (1998)}}
\title{Faddeev--type calculation of $\eta d$ threshold scattering}
\author{
N. V. Shevchenko$^{1,3}$, S. A. Rakityansky$^{1,2}$,
S. A. Sofianos$^{2}$, V. B. Belyaev$^{1,4}$\\ and W. Sandhas$^5$}
\address{
\parbox{14cm}{
\bigskip
{\small 1)}
Joint Institute  for Nuclear Research,Dubna, 141980, Russia\\
{\small 2)}
Physics Dept., University of South Africa,
P.O.Box 392, Pretoria, South Africa\\
{\small 3)}
Physics Dept., Irkutsk State University, Irkutsk 664003, Russia\\
{\small 4)}
	Research Center for Nuclear Physics, Osaka University, 
    Mihogaoko 10--1, Ibaraki, Osaka 567, Japan\\
{\small 5)}
Physikalisches Institut, Universit\H{a}t Bonn, D-53115 Bonn, Germany
}}
\maketitle
\begin{abstract}
The scattering length for the $\eta$-meson collision with deuteron is
calculated on the basis of rigorous few-body equations (AGS) for various
$\eta N$ input. The results obtained strongly support the existence of a
resonance or quasi-bound state close to the $\eta d$ threshold.\\

\end{abstract}
\bigskip
The production of $\eta$ mesons and their collisions with nuclei have been
studied experimentally and theoretically with increasing interest
during the last years. To a large extent this is motivated by the fundamental
questions of charge--symmetry breaking and the break--down of the
Okubo--Zweig--Iizuka rule. Another relevant question concerns the possible
formation of $\eta$--nucleus quasi--bound states.\\

In many respects the $\eta$--meson is similar to the $\pi^0$--meson despite
it's being four times heavier. Both are neutral and spinless, they have
almost
the same lifetime, $\sim 10^{-18}$ sec, and  are the only  mesons which have
a high probability of pure radiative decay, that is, their quarks can
annihilate into on-shell photons.
However, when being involved in nuclear reactions they behave rather
differently. The $S_{11}$--resonance $N^*(1535)$, for instance, is formed
in both $\pi N$ and $\eta N$ systems, but at different collision energies,
$$
\begin{array}{lcccrr}
	E^{res}_{\pi N}(S_{11}) &=& 1535\ {\rm MeV}\ - m_{N} -m_{\pi}
	&\approx& 458\ {\rm MeV}&\\
	\phantom{-}&&&&&\\
	E^{res}_{\eta N}(S_{11}) &=& 1535\ {\rm MeV}\ - m_{N} -m_{\eta}
	&\approx& 49\ {\rm MeV}\,.&\\
\end{array}
$$
Thus, due to the large mass of the $\eta$--meson (547.45\,MeV), this
resonance is very close to the $\eta N$--threshold. Furthermore it is  very
broad, with $\Gamma\approx 150\,{\rm MeV}$, covering the whole low energy
$\eta N$ region. As a result the interaction of nucleons with $\eta$--mesons
in this region, where the $S$--wave interaction dominates, is much stronger
than with pions.\\

After its creation the $N^*(1535)$--resonance decays into $\eta N$ and $\pi
N$ channels with equally high probabilities \cite{PDG}
\begin{equation}
\label{N*decay}
	N^*(1535)\rightarrow\left\{
	\begin{array}{ll}
	N+\eta & (35 - 55\ \%)\\
	N + \pi & (35 - 55\ \%)\\
	{\rm  other\ \ decays}&(\le10\ \%)\ \ \ \ ,\\
\end{array}
\right.
\end{equation}
which indicates that the $\eta N$ and $\pi N$ interactions are to be treated
by a coupled channel analysis. The resulting $\eta N$ interaction, obtained
in this way, turned out to be attractive \cite{bhal}. This raises the
question whether the attraction is strong enough to support $\eta$--nucleus
quasi--bound states. Let us recall in this context that, because of their
short lifetime, $\eta$--mesons can only be observed in final states of
certain nuclear reactions. Within nuclei they are considered to undergo
multiple absorption and production  processes via the $S_{11}$ resonance,
with a final transition into pions. Such quasi--bound states therefore would
be of considerable interest for studying $\eta$--meson properties in more
detail.\\

For the  calculation of these states various model treatments were employed,
among them the optical potential method \cite{ref3,ref4,ref5,ref6}, the
Green's function method \cite{ref7}, and the modified multiple scattering
theory \cite{gnw}. Calculations, based on  the exact
Alt--Grassberger--Sandhas (AGS) equations \cite{AGSeq},
for the $\eta d$ system were also made in Ref. \cite{ueda}
in early nineties.\\

The predictions concerning the possibility of $\eta$--mesic nucleus formation
are very diverse. One obvious reason for such a diversity is the poor
knowledge of the $\eta N$ forces. Another reason comes from the differences
among the employed approximations some of which might be detrimental in view
of the resonant character of the $\eta N$ dynamics and the delicacy of the
quasi--bound state problem. As was shown in Ref. \cite{ueda}
this problem  cannot be adequately addressed by a meson--nucleus
optical model or any low--order perturbation theory.\\

Among the approximate approaches the few--body dynamics of the
$\eta$--nucleus systems was most explicitly treated in our previous
calculations \cite{ours1,ours2,ours3,ours4,ours5} based on the finite--rank
approximation (FRA) of the nuclear Hamiltonians. The shortcoming of these
calculations is the neglect of excitations of the nuclear ground states. This
appears justified in the $\eta$$^4$He and possibly in the $\eta$--triton
($^3$He) case, but is quite questionable in $\eta$--deuteron collisions.\\

In the present paper we, therefore, treat the $\eta$--deuteron system on the
basis of the exact few--body equations (AGS). Both the $NN$ and $\eta N$
amplitudes entering them are chosen in separable form which reduces the
dimension of these equations to one. The same $\eta N$ amplitude have been
used in the FRA calculations. This allows us to compare our present
calculations of the $\eta d$ scattering length with the previous approximate
results, i.e., to examine the effect of the neglect of nuclear excitations
employed in the FRA. It turns out that the  discrepancies are not large for
most of the $\eta N$ parameter sets. This indicates that the conclusions
drawn in our previous investigation \cite{ours1,ours2,ours3,ours4,ours5} were
already fairly reliable and should be even more reliable in the less
sensitive $\eta$--triton or $\eta\,$$^4$He cases.\\

The $\eta$--deuteron scattering length is the value of the elastic
scattering amplitude
\begin{equation}
\label{ampl}
	  f({\bf p}_1',{\bf p}_1;z)=-(2\pi)^2\mu_1
	  \langle {\bf p}_1';\psi_d|U_{11}(z)|{\bf p}_1;\psi_d\rangle
\end{equation}
at zero collision energy. Here the subscript 1 labels the $\eta(NN)$
partition and the $\eta$--deuteron channel whose asymptotic states are
normalized as
$$
	  \langle {\bf p}_1';\psi_d|{\bf p}_1;\psi_d\rangle=
	  \delta({\bf p}_1'-{\bf p}_1)\ .
$$
The transition operator $U_{11}$ obeys the system of AGS equations
\begin{equation}
\label{ags}
     U_{ij}(z)=(1-\delta_{ij})g_0^{-1}(z)+\sum^3_{k=1}
     (1-\delta_{ik})t_k(z)g_0(z)U_{kj}(z)\ ,
     \qquad i,j=1,2,3\ ,
\end{equation}
where $g_0$ is the free Green's function in the three--body space, and
$t_i$ the two--body $T$--matrix for the $i$-th pair ($t_1=t_{NN}$).
For both $t_{\eta N}$ and $t_{NN}$ we used one--term separable forms
\begin{equation}
\label{sept}
       t_i(z)=|\chi_i\rangle\tau_i(z)\langle\chi_i|\ .
\end{equation}
For the $NN$ subsystem Eq. (\ref{sept}) implies that the asymptotic wave
function is related to the form-factor $|\chi_1\rangle$ according to
\begin{equation}
\label{dfun}
  |{\bf p}_1;\psi_d\rangle=g_0(z)|\chi_1\rangle|{\bf p}_1\rangle\ ,
\end{equation}
at $z=p_1^2/2\mu_1+E_d$ with $E_d$ being the deuteron energy.
Due  to (\ref{sept}) and
(\ref{dfun}) the scattering amplitude (\ref{ampl}) can be rewritten as
\begin{equation}
\label{ampl1}
	  f({\bf p}_1',{\bf p}_1;z)=-(2\pi)^2\mu_1
	  \langle {\bf p}_1'|X_{11}(z)|{\bf p}_1\rangle\ ,
\end{equation}
where the operators $X_{ij}$, defined as
$$
	   X_{ij}(z)=\langle\chi_i|g_0(z)U_{ij}(z)g_0(z)|\chi_j\rangle\ ,
$$
obey the system of equations
\begin{equation}
\label{agsx}
     X_{ij}(z)=Z_{ij}(z)+\sum^3_{k=1}Z_{ik}(z)
     \tau_k\left(z-\frac{p_k^2}{2\mu_k}\right)X_{kj}(z)
\end{equation}
with
$$
      Z_{ij}(z)=(1-\delta_{ij})\langle\chi_i|g_0(z)|\chi_j\rangle\ .
$$
The identity of the nucleons implies that $X_{31}=X_{21}$, $\tau_3=\tau_2$,
and $Z_{31}=Z_{21}$, which reduces the system (\ref{agsx}) to two coupled
equations
\begin{equation}
\label{agsi}
\left\{
\begin{array}{rcl}
  X_{11}(z) & = &
  \displaystyle
  2Z_{21}(z)\tau_2\left(z-\frac{p_2^2}{2\mu_2}\right)X_{21}(z)\ , \\
  &\\
  X_{21}(z) & = & Z_{21}(z)+
  \displaystyle
  Z_{21}(z)\tau_1\left(z-\frac{p_1^2}{2\mu_1}\right)X_{11}(z)+
  Z_{23}(z)\tau_2\left(z-\frac{p_2^2}{2\mu_2}\right)X_{21}(z)\ .\\
\end{array}\right.
\end{equation}
Eventually, after making the $S$--wave projection of the matrix elements
$\langle{\bf p}_i'|X_{ij}|{\bf p}_j\rangle$ and
$\langle{\bf p}_i'|Z_{ij}|{\bf p}_j\rangle$, we end up with
one--dimensional integral equations which can be solved numerically by
replacing the integrals by Gaussian sums.\\

The $S$--wave separable nucleon--nucleon and $\eta$--nucleon $T$--matrices of
the form (\ref{sept}) were adopted from Refs. \cite{garc} and \cite{ours1}.
However the parameters originally proposed in Ref.\cite{garc} for the
$T$--matrix
\begin{eqnarray*}
t_{NN}(p',p;z)&=&\frac{1}{4\pi}v(p')\frac{A(z)}{1-A(z)B(z)}v(p)\ ,\\
v(p)&=&\frac{\gamma}{\beta^2+p^2}\ ,\\
A(z)&=&-{\rm tgh}\left(1-\frac{z}{E_c}\right)\ ,\\
B(z)&=&\int_0^\infty\frac{p^2v^2(p)}{z-p^2/m_N+i\varepsilon}{\rm d}p\ ,
\end{eqnarray*}
were slightly modified to $E_c=0.816\,{\rm fm}^{-1}$,
$\beta=1.604\,{\rm fm}^{-1}$, and $\gamma^2=1.883\,{\rm fm}^{-2}$ which
correspond to more recent values of the triplet $NN$ scattering length,
$a_{NN}=5.424\,{\rm fm}$, and the effective range $r_{NN}=1.759\,{\rm fm}$
\cite{wong,dumbr}. With these parameters the deuteron is bound at 2.205\,MeV
and has an R.M.S.--radius $\sqrt{<r^2>_d}=1.887\,{\rm fm}$.\\

Instead of treating $\eta N$  and $\pi N$ as a two--channel system
it is customary  to describe the $\eta N$ interaction
by a one--channel complex potential. The strength  parameter $\lambda$ of
the corresponding $T$--matrix,
\begin{equation}
\label{tetan}
    t_{\eta N}(p',p;z)=
    \frac{\lambda}{({p'}^2+\alpha^2)(z-E_0+i\Gamma/2)(p^2+\alpha^2)}\ ,
\end{equation}
is chosen to reproduce the complex $\eta$--nucleon scattering length
$a_{\eta N}$,
\begin{equation}
\label{t000}
  \lambda=\frac{\alpha^4(E_0-i\Gamma/2)}{(2\pi)^2\mu_{\eta N}}a_{\eta N}\ ,
\end{equation}
the imaginary part of which accounts for the flux losses into the $\pi N$
channel. The range parameter $\alpha$ in (\ref{tetan}) is fixed in a
somewhat more complicated way (see Refs.\cite{bhal,bennh}), while $E_0$ and
$\Gamma$ are the parameters of the $S_{11}$ resonance \cite{PDG},
$$
E_0=1535\,{\rm MeV}-(m_N+m_\eta)\ ,\qquad \Gamma=150\,{\rm MeV}\ .
$$

The two--body scattering length $a_{\eta N}$, which defines the strength
parameter $\lambda$ via (\ref{t000}), is not accurately known. Different
analyses \cite{batinic} provided for $a_{\eta N}$ the values in the range
\begin{equation}
\label{interval}
0.27\ {\rm fm}\le{\rm Re\,}a_{\eta N}\le 0.98\ {\rm fm}\ ,\qquad
0.19\ {\rm fm}\le{\rm Im\,}a_{\eta N}\le 0.37\ {\rm fm}\ .
\end{equation}
The parameter $\alpha$ is also known with large uncertainty. Three
different values are given in the literature, namely,
$\alpha=2.357$ fm$^{-1}$ \cite{bhal}, $\alpha=3.316$ fm$^{-1}$ \cite{bennh},
and $\alpha=7.617$ fm$^{-1}$ \cite{bhal}.
We, therefore, calculate the $\eta$--deuteron scattering length $A_{\eta
d}$ for values of $a_{\eta N}$ and $\alpha$ covering these
intervals. The results of our calculations are given in Tables 2, and 3, and
also shown in Figs. 1 and 2.\\

In order to check our numerical procedure, we perform  test calculation
of the scattering length with decreasing values of the meson mass, and
compare the results obtained with the corresponding scattering lengths given
by
$$
A_{\eta d}^{\rm FSA}= -\frac{\mu}{\pi\alpha^4}\int_0^\infty
\left\{\frac{\mu}{\pi\alpha^4r}\left[r+\frac{e^{-\alpha r}}{2\alpha}
(3+\alpha r)-\frac{3}{2\alpha}\right]-\frac{1}{\lambda}(E_0-\frac{i}{2}
\Gamma)\right\}^{-1}\left|u_d(r)\right|^2\,{\rm d}r\ ,
$$
where $u_d$ is the radial wave function of the deuteron. This formula
is easily  derived in the Fixed Scatterer Approximation (FSA). As it
should be, the AGS and FSA results converge to each other
when the target particles become much heavier than the incident one
(see Table 1.) .\\

In Table 2 we compare the present AGS calculations with our previous results
obtained by means of the FRA \cite{ours1}. In Table 3 we present $A_{\eta d}$
calculated with $\alpha=3.316\,{\rm fm}^{-1}$ for various values of $a_{\eta
N}$ given in the literature. For comparison, we show also the results of
three different approximate calculations: that of Ref.\cite{gnw} where two
versions of the Multiple Scattering Theory (MST) were used, and a new
FRA--calculations which we performed with the deuteron wave function
(\ref{dfun}). In contrast to our previous FRA calculations this wave function
(which in the coordinate representation is of the Hulthen form) provides the
same $NN$ input as in the AGS calculations. The dependences of $A_{\eta d}$
on ${\rm Re\,}a_{\eta N}$ and ${\rm Im\,}a_{\eta N}$ are shown in Fig. 1 and
Fig. 2.\\

The curves depicted in Fig. 1 are similar to  Argand plots, though they
represent the scattering amplitude as a function of the coupling constant
instead of the collision energy. Despite this the circular
movement of the points on these curves is of the same nature as in the genuine
Argand plot. Indeed, to draw an Argand plot one moves the point on the
energy axis from left to right in the vicinity of a resonance pole. It is
clear that if we fix the energy instead and move the resonance pole itself
from right to left, the behavior of the amplitude should be similar to the
Argand circle. An increase of ${\rm Re\,}a_{\eta N}$ makes the $\eta N$
interaction more attractive which moves the $\eta d$ resonance poles
towards negative energies, i. e. from right to left. The Argand--like
shape of the curves in Fig. 1 implies therefore that at a certain value of
${\rm Re\,}a_{\eta N}$ (within the interval from 0.25\ fm  to 1\ fm)
the resonance pole bypasses (from below) the point $E=0$ and becomes
a quasi--bound pole.\\

It should be emphasized here that, in contrast to the genuine Argand plot,
all the points depicted in Fig. 1 correspond to the same energy, $E=0$,
and therefore the range in which the $\eta$-deuteron $S$--matrix pole moves
on the energy plane, when ${\rm Re\,}a_{\eta N}$ varies within its uncertainty
interval, cannot be inferred from these circular curves. They, however,
definitely indicate that such a pole exists and crosses the threshold line
${\rm Re\,}E=0$. The positions of this pole for different values of
${\rm Re\,}a_{\eta N}$ were explicitly calculated in a previous
publication \cite{ours5}, where the FRA--approximation was employed.
Recent measurements of $\eta$ production in the reaction
$p+n\rightarrow d+\eta$ show a substantial enhancement of the cross section
near threshold, as compared to what is expected from phase space analysis
\cite{sweden}, implying the existence of such a pole.\\

As can be seen in Table 3, both MST and FRA fail to give the correct
$A_{\eta d}$ (especially its real part) in the case of strong $\eta N$
interaction (when ${\rm Re\,}a_{\eta N}>0.5$\,fm) while for small values
of ${\rm Re\,}a_{\eta N}$ these methods work reasonably well.
Their failure in the case of strong two--body forces might be due to poor
convergence of the multiple scattering series and to  increased influence
of the break-up channel.\\

To summarize, in the present work we perform exact  AGS calculations  for
the $\eta d$ scattering length for various $\eta N$ input that include
new data which  appeared since the first calculation in 1991 \cite{ueda}.
The results obtained with these new
data strongly suggest that a resonance or quasi-bound state could
exist near the $\eta d$ threshold, in agreement with  the prediction
of Ref. \cite{ueda}.\\

\begin{center}
{\Large\bf ACKNOWLEDGMENTS}
\end{center}
Financial support from the Russian Foundation for Basic Research, Deutsche
Forschungsgemeinschaft, NATO (grant \#CR-GLG970110), INTAS (grant 
\#96-0457),
and the Foundation for Research Development of South
Africa is greatly appreciated.


\begin{table}
\caption{
Convergence of the AGS and FSA results for decreasing sequence of the
meson mass values. The parameters of the $\eta N$--potential are fixed by
$a_{\eta N}=(0.75+i0.27)$\,fm and $\alpha=2.357$\,fm$^{-1}$.
}
\begin{tabular}{|c|c|c|}
\hline
$\eta$--mass & $A_{\eta d}^{\rm AGS}$\,(fm) &
${A_{\eta d}^{\rm FSA}}_{\mathstrut}^{\mathstrut}$\,(fm)\\
\hline
$m_\eta$    & $3.941+i6.702$ & $1.936+i3.162$\\
$m_\eta/2 $ & $1.548+i0.596$ & $1.374+i0.856$\\
$m_\eta/3 $ & $0.891+i0.283$ & $0.878+i0.439$\\
$m_\eta/4 $ & $0.629+i0.185$ & $0.640+i0.292$\\
$m_\eta/5 $ & $0.487+i0.138$ & $0.503+i0.218$\\
$m_\eta/10$ & $0.230+i0.061$ & $0.242+i0.095$\\
$m_\eta/20$ & $0.113+i0.029$ & $0.119+i0.045$\\
$m_\eta/30$ & $0.075+i0.019$ & $0.079+i0.029$\\
$m_\eta/40$ & $0.056+i0.014$ & $0.059+i0.022$\\
$\phantom{-}m_\eta/50\phantom{-}$ &
$\phantom{-}0.045+i0.011\phantom{-}$ &
$\phantom{-}0.047+i0.017\phantom{-}$\\[0.1cm]
\hline
\end{tabular}
\end{table}
\begin{table}
\caption{
Comparison of $\eta d$ scattering lengths (in fm), obtained using the AGS 
and FRA methods, for 9 combinations of the parameters of the 
$\eta N$--potential.}
\begin{tabular}{|r|c|c|c|c|}
\hline
& \phantom{-}$\alpha=2.357\,({\rm fm}^{-1})^{\mathstrut}_{\mathstrut}
					   $\phantom{-} &
  \phantom{-}$\alpha=3.316\,({\rm fm}^{-1})$\phantom{-} &
  \phantom{-}$\alpha=7.617\,({\rm fm}^{-1})$\phantom{-} &
$a_{\eta N}\,({\rm fm})$\\
\hline
\parbox{1.5cm}{\begin{center}
AGS\\FRA
\end{center}} &
\parbox{3cm}{\begin{center}
$0.71+i0.79$\\ $0.66+i0.82$
\end{center}} &
\parbox{3cm}{\begin{center}
$0.71+i0.84$\\ $0.65+i0.85$
\end{center}} &
\parbox{3cm}{\begin{center}
$0.71+i0.92$\\ $0.62+i0.89$
\end{center}} &
\phantom{-}$0.27+i0.22$\phantom{-}\\
\hline
\parbox{1.5cm}{\begin{center}
AGS\\FRA
\end{center}} &
\parbox{3cm}{\begin{center}
$0.79+i0.68$\\ $0.75+i0.73$
\end{center}} &
\parbox{3cm}{\begin{center}
$0.81+i0.73$\\ $0.74+i0.76$
\end{center}} &
\parbox{3cm}{\begin{center}
$0.83+i0.81$\\ $0.72+i0.81$
\end{center}} &
\phantom{-}$0.28+i0.19$\phantom{-}\\
\hline
\parbox{1.5cm}{\begin{center}
AGS\\FRA
\end{center}} &
\parbox{3cm}{\begin{center}
$1.81+i2.44$\\ $1.53+i2.00$
\end{center}} &
\parbox{3cm}{\begin{center}
$1.64+i2.99$\\ $1.38+i2.15$
\end{center}} &
\parbox{3cm}{\begin{center}
$0.75+i4.00$\\ $1.14+i2.22$
\end{center}} &
\phantom{-}$0.55+i0.30$\phantom{-}\\
\hline
\end{tabular}
\end{table}
\begin{table}
\caption{
Results of the AGS and three different approximate calculations of
$A_{\eta d}$ with $\alpha=3.316\ {\rm fm}^{-1}$.
}
\begin{tabular}{|c|c|c||c|c|c|}
\hline
\multicolumn{2}{|c|}{$\eta N$--input} & exact
$A_{\eta d}^{\mathstrut}$\,(fm) &
\multicolumn{3}{c|}{approximate ${A_{\eta d}}_{\mathstrut}$\,(fm)}\\
\hline
\phantom{-}${\rm R^{\mathstrut}ef}_{\mathstrut}$.\phantom{-} &
$\phantom{-}a_{\eta N}$\,(fm)\phantom{-} &
\phantom{-}AGS\phantom{-} & MST I \cite{gnw} & MST II \cite{gnw}&
FRA\\
\hline
\cite{bennh} & $0.25\phantom{0}  + i0.16\phantom{0} $ & $0.73+i0.56$ &
   $0.66+i0.71$ & $0.66+i0.58$ & $0.65+i0.70$\\
\cite{bhal}  & $0.27\phantom{0}  + i0.22\phantom{0} $ & $0.71+i0.84$ &
   $0.57+i0.97$ & $0.64+i0.81$ & $0.59+i0.96$\\
\cite{krus}  & $\phantom{0}0.291 + i0.360\phantom{0}$ & $0.38+i1.36$ &
   $0.17+i1.35$ & $0.42+i1.25$ & $0.21+i1.35$\\
\cite{ref3}  & $0.30\phantom{0}  + i0.30\phantom{0} $ & $0.61+i1.22$ &
   $0.39+i1.28$ & $0.58+i1.11$ & $0.42+i1.27$\\
\cite{krus}  & $0.430            + i0.394           $ & $0.50+i2.07$ &
   $0.14+i1.91$ & $0.65+i1.73$ & $0.24+i1.88$\\
\cite{bhal}  & $0.44\phantom{0}  + i0.30\phantom{0} $ & $1.15+i1.89$ &
   $0.63+i1.93$ & $1.01+i1.50$ & $0.68+i1.86$\\
\cite{bennh} & $0.46\phantom{0}  + i0.29\phantom{0} $ & $1.31+i1.99$ &
   $0.72+i2.04$ & $1.11+i1.54$ & $0.76+i1.96$\\
\cite{bat1}  & $0.476            + i0.279           $ & $1.49+i2.06$ &
   $0.81+i2.15$ & $1.22+i1.56$ & $0.84+i2.05$\\
\cite{sauer} & $0.51\phantom{0}  + i0.21\phantom{0} $ & $2.37+i1.77$ &
   $1.48+i2.31$ & $1.65+i1.39$ & $1.38+i2.22$\\
\cite{ref3}  & $0.55\phantom{0}  + i0.30\phantom{0} $ & $1.64+i2.99$ &
   $0.61+i2.73$ & $1.40+i1.98$ & $0.69+i2.51$\\
\cite{krus}  & $0.579            + i0.399           $ & $0.34+i3.31$ &
   $-0.13+i2.64\phantom{0.}$ & $0.93+i2.41$ & $0.13+i2.52$\\
\cite{abaev} & $0.62\phantom{0}  + i0.30\phantom{0} $ & $1.80+i4.30$ &
   $0.36+i3.36$ & $1.65+i2.41$ & $0.55+i2.95$\\
\cite{bat1}  & $0.876            + i0.274           $ &
$-8.81+i4.30\phantom{0.}$ &
   $-2.76+i4.24\phantom{0.}$ & $2.42+i5.55$ & $-0.67+i3.98\phantom{0.}$\\
\cite{bat1}  & $0.888            + i0.274           $ &
$-8.63+i3.49\phantom{0.}$ &
   $-2.90+i4.12\phantom{0.}$ & $2.37+i5.79$ & $-0.73+i3.99\phantom{0.}$\\
\cite{arima} & $0.98\phantom{0}  + i0.37\phantom{0} $ &
$-4.69+i1.59\phantom{0.}$ &
   $-2.75+i2.77\phantom{0.}$ & $-0.06+i6.20\phantom{0.}$ &
					      $-1.18+i3.59\phantom{0.}$\\
\hline
\end{tabular}
\end{table}
\begin{figure}
\begin{center}
\unitlength=0.1mm
\begin{picture}(1300,1350)
\put(0,0){\line(1,0){1300}}
\put(0,1350){\line(1,0){1300}}
\put(0,0){\line(0,1){1350}}
\put(1300,0){\line(0,1){1350}}
\put(950,150){%
\begin{picture}(0,0)
\put(-900,0){\line(1,0){1200}}
\put(0,-100){\line(0,1){1200}}
\multiput(-900,0)(100,0){13}{\line(0,-1){20}}
\multiput(0,-100)(0,100){13}{\line(-1,0){20}}
\put(-935,-65){$-9$}
\put(-835,-65){$-8$}
\put(-735,-65){$-7$}
\put(-635,-65){$-6$}
\put(-535,-65){$-5$}
\put(-435,-65){$-4$}
\put(-235,-65){$-2$}
\put(-135,-65){$-1$}
\put(90,-65){1}
\put(190,-65){2}
\put(290,-65){3}
\put(-32,85){\llap{1}}
\put(-32,185){\llap{2}}
\put(-32,285){\llap{3}}
\put(-32,385){\llap{4}}
\put(-32,485){\llap{5}}
\put(-32,585){\llap{6}}
\put(-32,685){\llap{7}}
\put(-32,885){\llap{9}}
\put(-32,985){\llap{10}}
\put(-100,1130){${\rm Im\,}A_{\eta d}$ (fm)}
\put(300,30){\llap{${\rm Re\,}A_{\eta d}$ (fm)}}
\put(-565,905){\llap{2.357}}
\put(   047.937     ,  102.250   ){\circle*{10}}
\put(   051.032     ,  104.510   ){\circle*{10}}
\put(   054.194     ,  106.858   ){\circle*{10}}
\put(   057.427     ,  109.300   ){\circle*{10}}
\put(   060.731     ,  111.839   ){\circle*{10}}
\put(   064.111     ,  114.482   ){\circle*{10}}
\put(   067.568     ,  117.235   ){\circle*{10}}
\put(   071.105     ,  120.104   ){\circle*{10}}
\put(   074.725     ,  123.095   ){\circle*{10}}
\put(   078.429     ,  126.216   ){\circle*{10}}
\put(   082.222     ,  129.475   ){\circle*{10}}
\put(   086.106     ,  132.880   ){\circle*{10}}
\put(   090.084     ,  136.440   ){\circle*{10}}
\put(   094.159     ,  140.166   ){\circle*{10}}
\put(   098.333     ,  144.067   ){\circle*{10}}
\put(   102.611     ,  148.156   ){\circle*{10}}
\put(   106.994     ,  152.444   ){\circle*{10}}
\put(   111.487     ,  156.945   ){\circle*{10}}
\put(   116.091     ,  161.674   ){\circle*{10}}
\put(   120.811     ,  166.646   ){\circle*{10}}
\put(   125.648     ,  171.879   ){\circle*{10}}
\put(   130.605     ,  177.392   ){\circle*{10}}
\put(   135.684     ,  183.205   ){\circle*{10}}
\put(   140.888     ,  189.340   ){\circle*{10}}
\put(   146.218     ,  195.822   ){\circle*{10}}
\put(   151.675     ,  202.678   ){\circle*{10}}
\put(   157.259     ,  209.936   ){\circle*{10}}
\put(   162.968     ,  217.629   ){\circle*{10}}
\put(   168.802     ,  225.792   ){\circle*{10}}
\put(   174.757     ,  234.464   ){\circle*{10}}
\put(   180.828     ,  243.686   ){\circle*{10}}
\put(   187.007     ,  253.507   ){\circle*{10}}
\put(   193.283     ,  263.976   ){\circle*{10}}
\put(   199.645     ,  275.151   ){\circle*{10}}
\put(   206.074     ,  287.093   ){\circle*{10}}
\put(   212.546     ,  299.872   ){\circle*{10}}
\put(   219.034     ,  313.562   ){\circle*{10}}
\put(   225.501     ,  328.246   ){\circle*{10}}
\put(   231.900     ,  344.015   ){\circle*{10}}
\put(   238.175     ,  360.969   ){\circle*{10}}
\put(   244.254     ,  379.215   ){\circle*{10}}
\put(   250.047     ,  398.872   ){\circle*{10}}
\put(   255.446     ,  420.065   ){\circle*{10}}
\put(   260.316     ,  442.931   ){\circle*{10}}
\put(   264.490     ,  467.612   ){\circle*{10}}
\put(   267.765     ,  494.255   ){\circle*{10}}
\put(   269.894     ,  523.010   ){\circle*{10}}
\put(   270.573     ,  554.021   ){\circle*{10}}
\put(   269.440     ,  587.418   ){\circle*{10}}
\put(   266.055     ,  623.308   ){\circle*{10}}
\put(   259.898     ,  661.750   ){\circle*{10}}
\put(   250.357     ,  702.740   ){\circle*{10}}
\put(   236.725     ,  746.172   ){\circle*{10}}
\put(   218.199     ,  791.803   ){\circle*{10}}
\put(   193.901     ,  839.199   ){\circle*{10}}
\put(   162.905     ,  887.684   ){\circle*{10}}
\put(   124.297     ,  936.281   ){\circle*{10}}
\put(   077.268     ,  983.663   ){\circle*{10}}
\put(   021.235     , 1028.130   ){\circle*{10}}
\put(  -043.998     , 1067.631   ){\circle*{10}}
\put(  -118.071     , 1099.857   ){\circle*{10}}
\put(  -199.914     , 1122.417   ){\circle*{10}}
\put(  -287.666     , 1133.104   ){\circle*{10}}
\put(  -378.711     , 1130.212   ){\circle*{10}}
\put(  -469.862     , 1112.839   ){\circle*{10}}
\put(  -557.691     , 1081.104   ){\circle*{10}}
\put(  -638.933     , 1036.188   ){\circle*{10}}
\put(  -710.883     ,  980.193   ){\circle*{10}}
\put(  -771.674     ,  915.843   ){\circle*{10}}
\put(  -820.388     ,  846.113   ){\circle*{10}}
\put(  -857.005     ,  773.874   ){\circle*{10}}
\put(  -882.225     ,  701.636   ){\circle*{10}}
\put(  -897.235     ,  631.392   ){\circle*{10}}
\put(  -903.483     ,  564.589   ){\circle*{10}}
\put(  -902.489     ,  502.159   ){\circle*{10}}
\put(  -895.709     ,  444.608   ){\circle*{10}}
\put(-450,700){\llap{3.316}}
\put(  044.856     ,   106.935  ){\circle*{10}}
\put(  047.865     ,   109.621  ){\circle*{10}}
\put(  050.940     ,   112.427  ){\circle*{10}}
\put(  054.083     ,   115.361  ){\circle*{10}}
\put(  057.294     ,   118.431  ){\circle*{10}}
\put(  060.575     ,   121.645  ){\circle*{10}}
\put(  063.930     ,   125.012  ){\circle*{10}}
\put(  067.358     ,   128.544  ){\circle*{10}}
\put(  070.863     ,   132.250  ){\circle*{10}}
\put(  074.445     ,   136.143  ){\circle*{10}}
\put(  078.106     ,   140.235  ){\circle*{10}}
\put(  081.848     ,   144.542  ){\circle*{10}}
\put(  085.670     ,   149.077  ){\circle*{10}}
\put(  089.574     ,   153.857  ){\circle*{10}}
\put(  093.560     ,   158.902  ){\circle*{10}}
\put(  097.627     ,   164.230  ){\circle*{10}}
\put(  101.776     ,   169.863  ){\circle*{10}}
\put(  106.003     ,   175.825  ){\circle*{10}}
\put(  110.307     ,   182.142  ){\circle*{10}}
\put(  114.684     ,   188.842  ){\circle*{10}}
\put(  119.128     ,   195.956  ){\circle*{10}}
\put(  123.633     ,   203.519  ){\circle*{10}}
\put(  128.189     ,   211.568  ){\circle*{10}}
\put(  132.785     ,   220.143  ){\circle*{10}}
\put(  137.405     ,   229.291  ){\circle*{10}}
\put(  142.033     ,   239.060  ){\circle*{10}}
\put(  146.643     ,   249.506  ){\circle*{10}}
\put(  151.206     ,   260.687  ){\circle*{10}}
\put(  155.688     ,   272.669  ){\circle*{10}}
\put(  160.043     ,   285.523  ){\circle*{10}}
\put(  164.217     ,   299.326  ){\circle*{10}}
\put(  168.142     ,   314.163  ){\circle*{10}}
\put(  171.737     ,   330.122  ){\circle*{10}}
\put(  174.899     ,   347.300  ){\circle*{10}}
\put(  177.507     ,   365.797  ){\circle*{10}}
\put(  179.411     ,   385.717  ){\circle*{10}}
\put(  180.428     ,   407.165  ){\circle*{10}}
\put(  180.340     ,   430.241  ){\circle*{10}}
\put(  178.881     ,   455.036  ){\circle*{10}}
\put(  175.735     ,   481.623  ){\circle*{10}}
\put(  170.527     ,   510.045  ){\circle*{10}}
\put(  162.817     ,   540.298  ){\circle*{10}}
\put(  152.093     ,   572.308  ){\circle*{10}}
\put(  137.780     ,   605.906  ){\circle*{10}}
\put(  119.239     ,   640.789  ){\circle*{10}}
\put(  095.793     ,   676.486  ){\circle*{10}}
\put(  066.765     ,   712.311  ){\circle*{10}}
\put(  031.536     ,   747.327  ){\circle*{10}}
\put( -010.368     ,   780.322  ){\circle*{10}}
\put( -059.162     ,   809.812  ){\circle*{10}}
\put( -114.680     ,   834.097  ){\circle*{10}}
\put( -176.248     ,   851.371  ){\circle*{10}}
\put( -242.604     ,   859.904  ){\circle*{10}}
\put( -311.892     ,   858.277  ){\circle*{10}}
\put( -381.766     ,   845.620  ){\circle*{10}}
\put( -449.607     ,   821.807  ){\circle*{10}}
\put( -512.828     ,   787.538  ){\circle*{10}}
\put( -569.188     ,   744.276  ){\circle*{10}}
\put( -617.044     ,   694.045  ){\circle*{10}}
\put( -655.487     ,   639.157  ){\circle*{10}}
\put( -684.336     ,   581.916  ){\circle*{10}}
\put( -704.025     ,   524.390  ){\circle*{10}}
\put( -715.429     ,   468.263  ){\circle*{10}}
\put( -719.675     ,   414.781  ){\circle*{10}}
\put( -717.982     ,   364.767  ){\circle*{10}}
\put( -711.540     ,   318.682  ){\circle*{10}}
\put( -701.435     ,   276.702  ){\circle*{10}}
\put( -688.612     ,   238.793  ){\circle*{10}}
\put( -673.859     ,   204.785  ){\circle*{10}}
\put( -657.817     ,   174.420  ){\circle*{10}}
\put( -640.995     ,   147.400  ){\circle*{10}}
\put( -623.786     ,   123.409  ){\circle*{10}}
\put( -606.487     ,   102.135  ){\circle*{10}}
\put( -589.321     ,   083.281  ){\circle*{10}}
\put( -572.448     ,   066.573  ){\circle*{10}}
\put( -555.982     ,   051.761  ){\circle*{10}}
\put(-320,470){\llap{7.617}}
\put(    038.904   ,   112.951   ){\circle*{10}}
\put(    041.601   ,   116.303   ){\circle*{10}}
\put(    044.339   ,   119.834   ){\circle*{10}}
\put(    047.119   ,   123.559   ){\circle*{10}}
\put(    049.938   ,   127.491   ){\circle*{10}}
\put(    052.795   ,   131.646   ){\circle*{10}}
\put(    055.687   ,   136.041   ){\circle*{10}}
\put(    058.609   ,   140.695   ){\circle*{10}}
\put(    061.558   ,   145.628   ){\circle*{10}}
\put(    064.527   ,   150.863   ){\circle*{10}}
\put(    067.508   ,   156.425   ){\circle*{10}}
\put(    070.493   ,   162.340   ){\circle*{10}}
\put(    073.469   ,   168.638   ){\circle*{10}}
\put(    076.423   ,   175.351   ){\circle*{10}}
\put(    079.337   ,   182.515   ){\circle*{10}}
\put(    082.190   ,   190.168   ){\circle*{10}}
\put(    084.957   ,   198.353   ){\circle*{10}}
\put(    087.604   ,   207.115   ){\circle*{10}}
\put(    090.094   ,   216.506   ){\circle*{10}}
\put(    092.380   ,   226.579   ){\circle*{10}}
\put(    094.406   ,   237.393   ){\circle*{10}}
\put(    096.100   ,   249.009   ){\circle*{10}}
\put(    097.381   ,   261.492   ){\circle*{10}}
\put(    098.145   ,   274.910   ){\circle*{10}}
\put(    098.269   ,   289.331   ){\circle*{10}}
\put(    097.606   ,   304.821   ){\circle*{10}}
\put(    095.977   ,   321.439   ){\circle*{10}}
\put(    093.167   ,   339.234   ){\circle*{10}}
\put(    088.922   ,   358.235   ){\circle*{10}}
\put(    082.944   ,   378.442   ){\circle*{10}}
\put(    074.884   ,   399.808   ){\circle*{10}}
\put(    064.343   ,   422.226   ){\circle*{10}}
\put(    050.878   ,   445.498   ){\circle*{10}}
\put(    034.008   ,   469.311   ){\circle*{10}}
\put(    013.243   ,   493.203   ){\circle*{10}}
\put(   -011.880   ,   516.532   ){\circle*{10}}
\put(   -041.740   ,   538.450   ){\circle*{10}}
\put(   -076.545   ,   557.899   ){\circle*{10}}
\put(   -116.240   ,   573.631   ){\circle*{10}}
\put(   -160.402   ,   584.286   ){\circle*{10}}
\put(   -208.153   ,   588.516   ){\circle*{10}}
\put(   -258.138   ,   585.167   ){\circle*{10}}
\put(   -308.574   ,   573.483   ){\circle*{10}}
\put(   -357.412   ,   553.286   ){\circle*{10}}
\put(   -402.576   ,   525.071   ){\circle*{10}}
\put(   -442.243   ,   489.985   ){\circle*{10}}
\put(   -475.069   ,   449.670   ){\circle*{10}}
\put(   -500.331   ,   406.028   ){\circle*{10}}
\put(   -517.930   ,   360.961   ){\circle*{10}}
\put(   -528.295   ,   316.160   ){\circle*{10}}
\put(   -532.225   ,   272.973   ){\circle*{10}}
\put(   -530.722   ,   232.361   ){\circle*{10}}
\put(   -524.840   ,   194.915   ){\circle*{10}}
\put(   -515.586   ,   160.919   ){\circle*{10}}
\put(   -503.856   ,   130.422   ){\circle*{10}}
\put(   -490.406   ,   103.312   ){\circle*{10}}
\put(   -475.851   ,   079.372   ){\circle*{10}}
\put(   -460.674   ,   058.336   ){\circle*{10}}
\put(   -445.243   ,   039.911   ){\circle*{10}}
\put(   -429.831   ,   023.808   ){\circle*{10}}
\put(   -414.636   ,   009.750   ){\circle*{10}}
\put(   -399.796   ,  -002.516   ){\circle*{10}}
\put(   -385.403   ,  -013.221   ){\circle*{10}}
\put(   -371.513   ,  -022.569   ){\circle*{10}}
\put(   -358.159   ,  -030.740   ){\circle*{10}}
\put(   -345.354   ,  -037.891   ){\circle*{10}}
\put(   -333.098   ,  -044.159   ){\circle*{10}}
\put(   -321.380   ,  -049.661   ){\circle*{10}}
\put(   -310.185   ,  -054.500   ){\circle*{10}}
\put(   -299.492   ,  -058.764   ){\circle*{10}}
\put(   -289.280   ,  -062.529   ){\circle*{10}}
\put(   -279.523   ,  -065.861   ){\circle*{10}}
\put(   -270.198   ,  -068.817   ){\circle*{10}}
\put(   -261.280   ,  -071.445   ){\circle*{10}}
\put(   -252.745   ,  -073.788   ){\circle*{10}}
\put(   -244.570   ,  -075.883   ){\circle*{10}}
\end{picture}}
\end{picture}
\end{center}
\caption{
The values of $A_{\eta d}$ calculated for ${\rm Im\,}a_{\eta N}=0.30$\,fm
while ${\rm Re\,}a_{\eta N}$ is changing from 0.25\,fm to 1\,fm with the
step 0.01\,fm. An increase of ${\rm Re\,}a_{\eta N}$ moves the points in the
anti-clockwise direction along the curve trajectories which correspond to
three choices of the range parameter $\alpha$.
}
\end{figure}
\begin{figure}
\begin{center}
\unitlength=0.1mm
\begin{picture}(1000,900)
\put(0,0){\line(1,0){1000}}
\put(0,900){\line(1,0){1000}}
\put(0,0){\line(0,1){900}}
\put(1000,0){\line(0,1){900}}
\put(250,100){%
\begin{picture}(0,0)
\put(-200,0){\line(1,0){900}}
\put(0,0){\line(0,1){700}}
\multiput(-200,0)(100,0){10}{\line(0,-1){20}}
\multiput(0,0)(0,100){8}{\line(-1,0){20}}
\put(-235,-65){$-2$}
\put(-135,-65){$-1$}
\put(-10,-65){0}
\put(90,-65){1}
\put(190,-65){2}
\put(290,-65){3}
\put(390,-65){4}
\put(490,-65){5}
\put(590,-65){6}
\put(690,-65){7}
\put(-32,85){\llap{1}}
\put(-32,185){\llap{2}}
\put(-32,285){\llap{3}}
\put(-32,385){\llap{4}}
\put(-32,585){\llap{6}}
\put(-100,730){${\rm Im\,}A_{\eta d}$ (fm)}
\put(700,30){\llap{${\rm Re\,}A_{\eta d}$ (fm)}}
\put(340,130){\llap{2.357}}
\put(  328.597   ,   188.334   ){\circle*{10}}
\put(  324.296   ,   195.960   ){\circle*{10}}
\put(  319.765   ,   203.410   ){\circle*{10}}
\put(  315.016   ,   210.674   ){\circle*{10}}
\put(  310.060   ,   217.741   ){\circle*{10}}
\put(  304.908   ,   224.604   ){\circle*{10}}
\put(  299.572   ,   231.255   ){\circle*{10}}
\put(  294.064   ,   237.687   ){\circle*{10}}
\put(  288.396   ,   243.894   ){\circle*{10}}
\put(  282.582   ,   249.872   ){\circle*{10}}
\put(  276.634   ,   255.615   ){\circle*{10}}
\put(  270.565   ,   261.122   ){\circle*{10}}
\put(  264.388   ,   266.389   ){\circle*{10}}
\put(  258.115   ,   271.414   ){\circle*{10}}
\put(  251.759   ,   276.198   ){\circle*{10}}
\put(  245.332   ,   280.740   ){\circle*{10}}
\put(  238.847   ,   285.040   ){\circle*{10}}
\put(  232.314   ,   289.102   ){\circle*{10}}
\put(  225.746   ,   292.925   ){\circle*{10}}
\put(  219.153   ,   296.514   ){\circle*{10}}
\put(  212.546   ,   299.872   ){\circle*{10}}
\put(  205.935   ,   303.001   ){\circle*{10}}
\put(  199.330   ,   305.908   ){\circle*{10}}
\put(  192.739   ,   308.595   ){\circle*{10}}
\put(  186.172   ,   311.069   ){\circle*{10}}
\put(  179.636   ,   313.335   ){\circle*{10}}
\put(  173.140   ,   315.398   ){\circle*{10}}
\put(  166.689   ,   317.264   ){\circle*{10}}
\put(  160.291   ,   318.940   ){\circle*{10}}
\put(  153.952   ,   320.432   ){\circle*{10}}
\put(  147.678   ,   321.745   ){\circle*{10}}
\put(  141.473   ,   322.887   ){\circle*{10}}
\put(  135.342   ,   323.863   ){\circle*{10}}
\put(  129.290   ,   324.681   ){\circle*{10}}
\put(  123.319   ,   325.347   ){\circle*{10}}
\put(  117.434   ,   325.868   ){\circle*{10}}
\put(  111.638   ,   326.249   ){\circle*{10}}
\put(  105.932   ,   326.497   ){\circle*{10}}
\put(  100.319   ,   326.619   ){\circle*{10}}
\put(  094.801   ,   326.620   ){\circle*{10}}
\put(  089.380   ,   326.507   ){\circle*{10}}
\put(500,215){\llap{3.316}}
\put(   395.423    ,   267.576   ){\circle*{10}}
\put(   386.427    ,   278.954   ){\circle*{10}}
\put(   376.974    ,   289.793   ){\circle*{10}}
\put(   367.110    ,   300.068   ){\circle*{10}}
\put(   356.882    ,   309.759   ){\circle*{10}}
\put(   346.337    ,   318.851   ){\circle*{10}}
\put(   335.525    ,   327.334   ){\circle*{10}}
\put(   324.492    ,   335.203   ){\circle*{10}}
\put(   313.285    ,   342.455   ){\circle*{10}}
\put(   301.949    ,   349.096   ){\circle*{10}}
\put(   290.528    ,   355.131   ){\circle*{10}}
\put(   279.063    ,   360.571   ){\circle*{10}}
\put(   267.593    ,   365.429   ){\circle*{10}}
\put(   256.154    ,   369.723   ){\circle*{10}}
\put(   244.779    ,   373.470   ){\circle*{10}}
\put(   233.500    ,   376.692   ){\circle*{10}}
\put(   222.345    ,   379.409   ){\circle*{10}}
\put(   211.337    ,   381.646   ){\circle*{10}}
\put(   200.500    ,   383.427   ){\circle*{10}}
\put(   189.852    ,   384.776   ){\circle*{10}}
\put(   179.411    ,   385.717   ){\circle*{10}}
\put(   169.190    ,   386.277   ){\circle*{10}}
\put(   159.201    ,   386.480   ){\circle*{10}}
\put(   149.455    ,   386.350   ){\circle*{10}}
\put(   139.958    ,   385.911   ){\circle*{10}}
\put(   130.716    ,   385.187   ){\circle*{10}}
\put(   121.734    ,   384.198   ){\circle*{10}}
\put(   113.013    ,   382.968   ){\circle*{10}}
\put(   104.555    ,   381.517   ){\circle*{10}}
\put(   096.359    ,   379.863   ){\circle*{10}}
\put(   088.425    ,   378.027   ){\circle*{10}}
\put(   080.750    ,   376.025   ){\circle*{10}}
\put(   073.330    ,   373.875   ){\circle*{10}}
\put(   066.162    ,   371.591   ){\circle*{10}}
\put(   059.242    ,   369.189   ){\circle*{10}}
\put(   052.565    ,   366.682   ){\circle*{10}}
\put(   046.125    ,   364.084   ){\circle*{10}}
\put(   039.916    ,   361.405   ){\circle*{10}}
\put(   033.934    ,   358.657   ){\circle*{10}}
\put(   028.171    ,   355.850   ){\circle*{10}}
\put(   022.621    ,   352.994   ){\circle*{10}}
\put(640,575){\llap{7.617}}
\put(  599.407  ,   614.832  ){\circle*{10}}
\put(  553.300  ,   639.760  ){\circle*{10}}
\put(  506.456  ,   658.898  ){\circle*{10}}
\put(  459.794  ,   672.514  ){\circle*{10}}
\put(  414.093  ,   681.028  ){\circle*{10}}
\put(  369.977  ,   684.957  ){\circle*{10}}
\put(  327.915  ,   684.874  ){\circle*{10}}
\put(  288.228  ,   681.363  ){\circle*{10}}
\put(  251.114  ,   674.992  ){\circle*{10}}
\put(  216.661  ,   666.294  ){\circle*{10}}
\put(  184.875  ,   655.748  ){\circle*{10}}
\put(  155.700  ,   643.780  ){\circle*{10}}
\put(  129.032  ,   630.757  ){\circle*{10}}
\put(  104.740  ,   616.990  ){\circle*{10}}
\put(  082.671  ,   602.739  ){\circle*{10}}
\put(  062.667  ,   588.218  ){\circle*{10}}
\put(  044.566  ,   573.602  ){\circle*{10}}
\put(  028.209  ,   559.028  ){\circle*{10}}
\put(  013.443  ,   544.605  ){\circle*{10}}
\put(  000.125  ,   530.419  ){\circle*{10}}
\put( -011.880  ,   516.532  ){\circle*{10}}
\put( -022.697  ,   502.992  ){\circle*{10}}
\put( -032.440  ,   489.833  ){\circle*{10}}
\put( -041.214  ,   477.077  ){\circle*{10}}
\put( -049.112  ,   464.737  ){\circle*{10}}
\put( -056.222  ,   452.820  ){\circle*{10}}
\put( -062.620  ,   441.327  ){\circle*{10}}
\put( -068.376  ,   430.256  ){\circle*{10}}
\put( -073.554  ,   419.599  ){\circle*{10}}
\put( -078.210  ,   409.349  ){\circle*{10}}
\put( -082.395  ,   399.494  ){\circle*{10}}
\put( -086.154  ,   390.023  ){\circle*{10}}
\put( -089.529  ,   380.923  ){\circle*{10}}
\put( -092.557  ,   372.181  ){\circle*{10}}
\put( -095.271  ,   363.784  ){\circle*{10}}
\put( -097.701  ,   355.718  ){\circle*{10}}
\put( -099.873  ,   347.970  ){\circle*{10}}
\put( -101.812  ,   340.526  ){\circle*{10}}
\put( -103.539  ,   333.373  ){\circle*{10}}
\put( -105.074  ,   326.500  ){\circle*{10}}
\put( -106.435  ,   319.894  ){\circle*{10}}
\end{picture}}
\end{picture}
\end{center}
\caption{
The values of $A_{\eta d}$ calculated for ${\rm Re\,}a_{\eta N}=0.60$\,fm
while ${\rm Im\,}a_{\eta N}$ is changing from 0.2\,fm to 0.4\,fm with the
step 0.005\,fm. An increase of ${\rm Im\,}a_{\eta N}$ moves the points in the
anti-clockwise direction along the curve trajectories which correspond to
three choices of the range parameter $\alpha$.
}
\end{figure}
\end{document}